# Distributions of the charged particles multiplicities in the electromagnetic showers initiated by 10 to 1000 GeV electrons in lead

S.P. Denisov, V.N. Goryachev

Institute for High Energy Physics of the National Research Centre «Kurchatov Institute»
1 Nauki sqr, Protvino, Moscow region, 142281 Russian Federation

**Abstract**

Distributions of the charged particles multiplicities in the electromagnetic showers initiated by 10 to 1000 GeV electrons in lead are calculated using GEANT4. It is shown that they are well fitted by the inverse sum of two exponents. The evolution of the multiplicity distribution shapes as a function of the lead depth is discussed. An estimate of the energy resolution of a simple *e,γ* detector consisting of a high *Z* convertor and a counter of the shower electrons and positrons is presented.

## 1. Introduction

Detectors consisting of a lead converter and a position sensitive detector behind it are widely used in HEP experiments for *e,γ*/hadron  and $\gamma/\pi^0$ separations and for *e, γ* coordinate and energy measurements[1-13]. They are often called preshower or shower maximum detectors. Their characteristics depend on the fluctuations of the charge particle flux in electromagnetic (EM) showers initiated in the converter by the primary *e,γ*. In our previous studies[14] it is shown that multiplicity distributions at the converter depth of $t_{\max}$, corresponding to the maximum of the charged particles flux in the EM showers, are asymmetric and have long tails at the low multiplicities due to late developments of some showers. They are well described by the inverse sum of two exponents:

$$\frac{dP}{dN} = \frac{p_0}{e^{p_1(N-p_3)} + e^{p_2(N-p_3)}}, \qquad p_0 = \frac{p_1 - p_2}{\pi} \sin \frac{\pi p_1}{p_1 - p_2} \tag{1}$$

where $p_0$ is a normalization factor and $p_1$, $p_2$, $p_3$ are free parameters. This function is defined from $-\infty$ to $+\infty$ and because $N>0$ it can be used only if the multiplicity $N>>1$ and $dP/dN(0)$ is close to 0. Thus, it is applicable only to thick converters with $t>>1$ (here and below $t$ is in radiation lengths). The main purpose of this report is to study how well this formula works at the converter thickness $t$ other than $t_{max}$ and how $dP/dN$ shape varies as a function of $t$.

To answer these questions 5000 EM cascades in lead were generated for each primary electron energy $E_0$ of 10, 20, 40, 80, 160, 200, 500 and 1000 GeV. The calculations were based on GEANT4 10.01.p02 (Physical list FTFP_BERT)[15] with 700 micron range cut. Corresponding energy thresholds for the secondaries $e^+$ and $e^-$ are close to 1 MeV. From Table 1 it follows that increasing or decreasing the range cut by a factor of two does not affect the average particle flux $<N>$ within statistical uncertainty of 0.5% since the energy thresholds are much less than the average particle energy[16]. Note that the charged particles flux consists mainly of $e^+$ and $e^-$. For example the admixture of other particles at $t_{max}$ for $E_0=200$ GeV is 0.02% only[14]. The converter diameter is 70 cm.

Table 1. $<N>$ vs range cut

| $E_0$, GeV | $t$, r. l. | Cut, mm | | |
|---|---|---|---|---|
| | | 0.35 | 0.7 | 1.4 |
| 40 | 16 | 36.4 ± 0.2 | 36.2 ± 0.2 | 36.1 ± 0.2 |
| 80 | 16 | 93.7 ± 0.6 | 90.5 ± 0.6 | 90.9 ± 0.6 |
| 200 | 9 | 828.1 ± 3.5 | 825.9 ± 3.5 | 825.6 ± 3.5 |

**2. Charge particle multiplicity distributions**

Fig.1 shows the results of GEANT4 simulations of EM showers in lead fitted to gamma function

$$<N> = c_0(bt)^{a-1}e^{-bt}, \tag{2}$$

where $c_0$, $a$ and $b$ are free parameters[17].

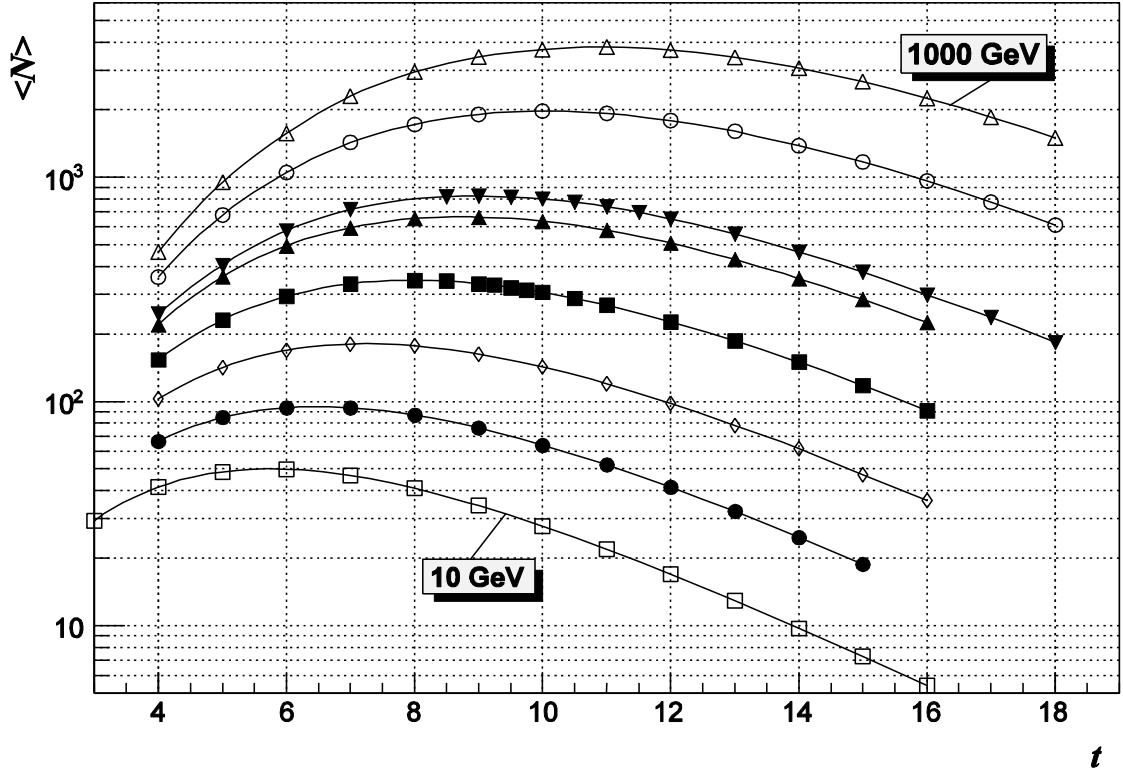

Figure 1. Results of GEANT4 simulation of EM showers in lead fitted to gamma function (2).

$dP/dN$ distributions for $E_0$=10, 40, 200 and 1000 GeV are presented in Figs.2 - 5. For all energies and $t \gg 1$ these distributions are reasonably well described by formula (1). Fig.6 presents the $t$-dependencies of $p_1$ and $-p_2$ parameters obtained from the fit to formula (1). From Fig.6 it follows that at some depths $t_1$ below and $t_2$ above $t_{max}$ $p_1$ and $-p_2$ parameters become equal (two $p_1$, $-p_2$ intersections below $t_{max}$ at 1000 GeV are not discussed in this paper). As follows from formula (1) at these $t$ values $dP/dN$ distributions become symmetric (see Figs. 2 - 5) and can be described by the following function:

$$dP/dN = (p/\pi) \cdot \text{ch}^{-1}(p[N-p_3]), \qquad (3)$$

where $p = p_1 = -p_2$. This function reaches maximum of $p/\pi$ at $N = p_3$. Asymmetric $dP/dN$ distributions between $t_1$ and $t_2$ have tails on the left side of the $dP/dN$ peak while in the region of $t > t_2$ the tail appears on the right side as shown in Fig. 2 - 5.

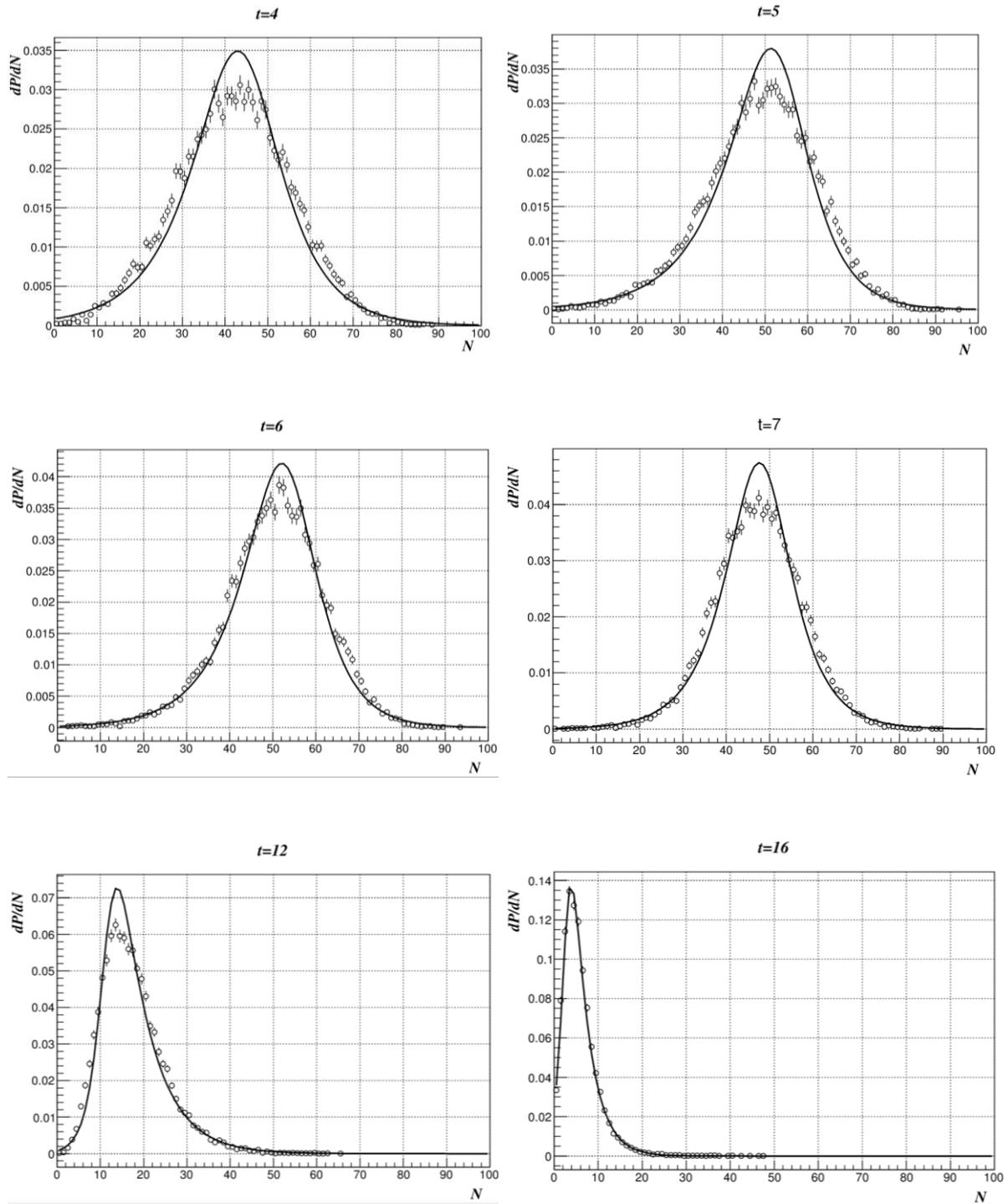

Figure 2. Charged particles multiplicity distributions in the EM showers initiated by 10 GeV electrons at different depths *t* of lead.

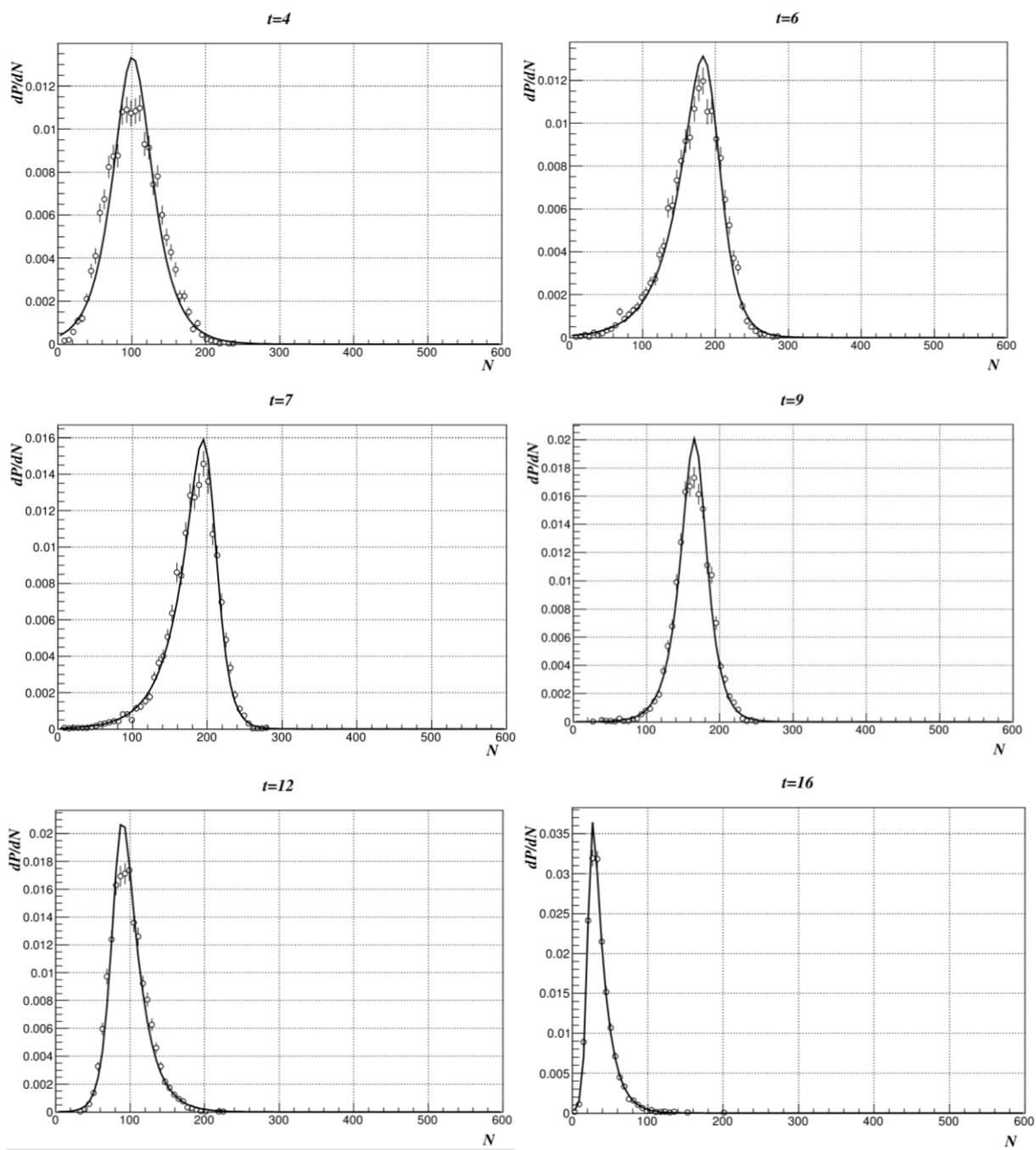

Figure 3. Charged particles multiplicity distributions in the EM showers initiated by 40 GeV electrons at different depths $t$ of lead.

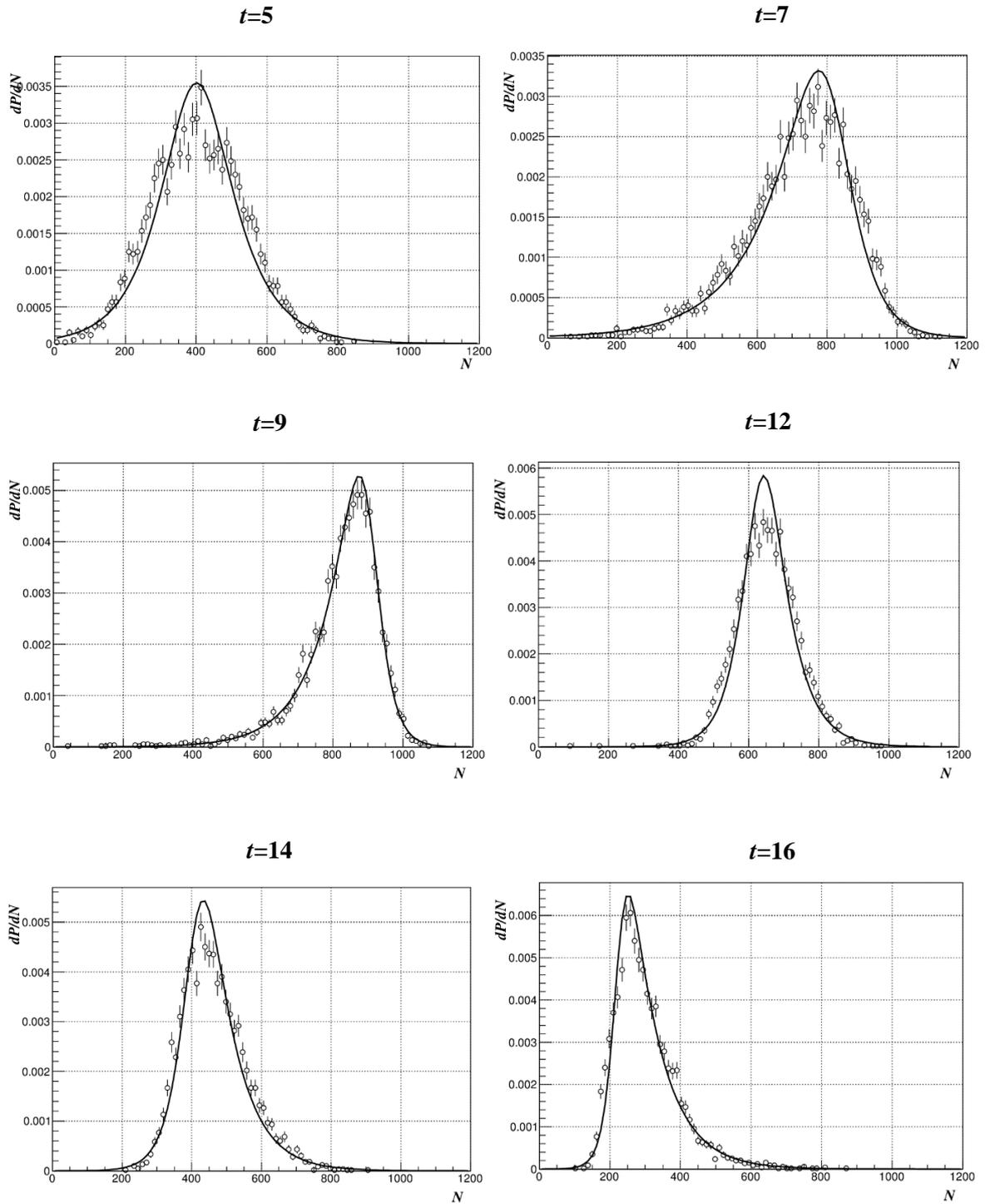

Figure 4. Charged particles multiplicity distributions in the EM showers initiated by 200 GeV electrons at different depths *t* of lead.

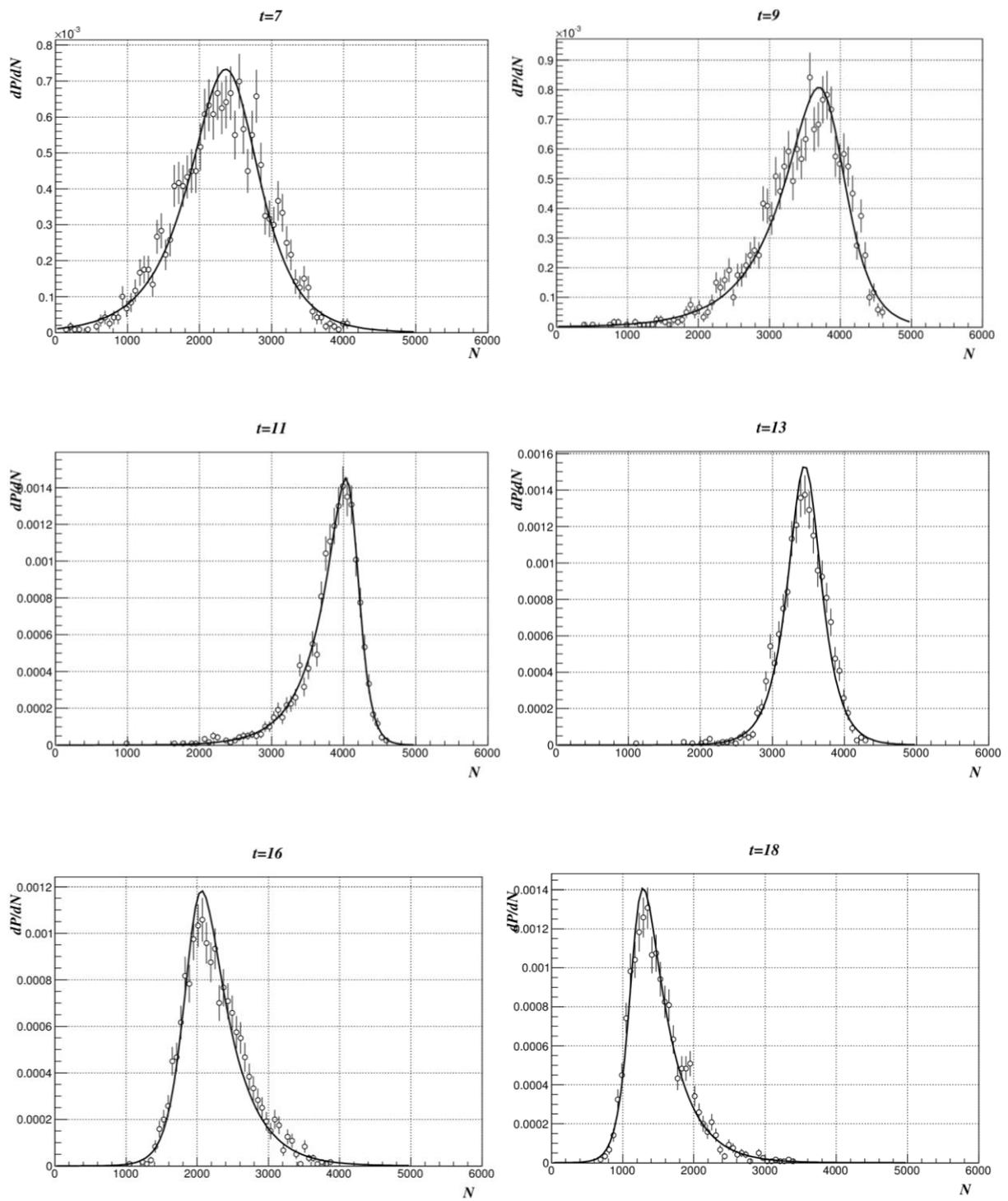

Figure 5. Charged particles multiplicity distributions in the EM showers initiated by 1000 GeV electrons at different depths *t* of lead.

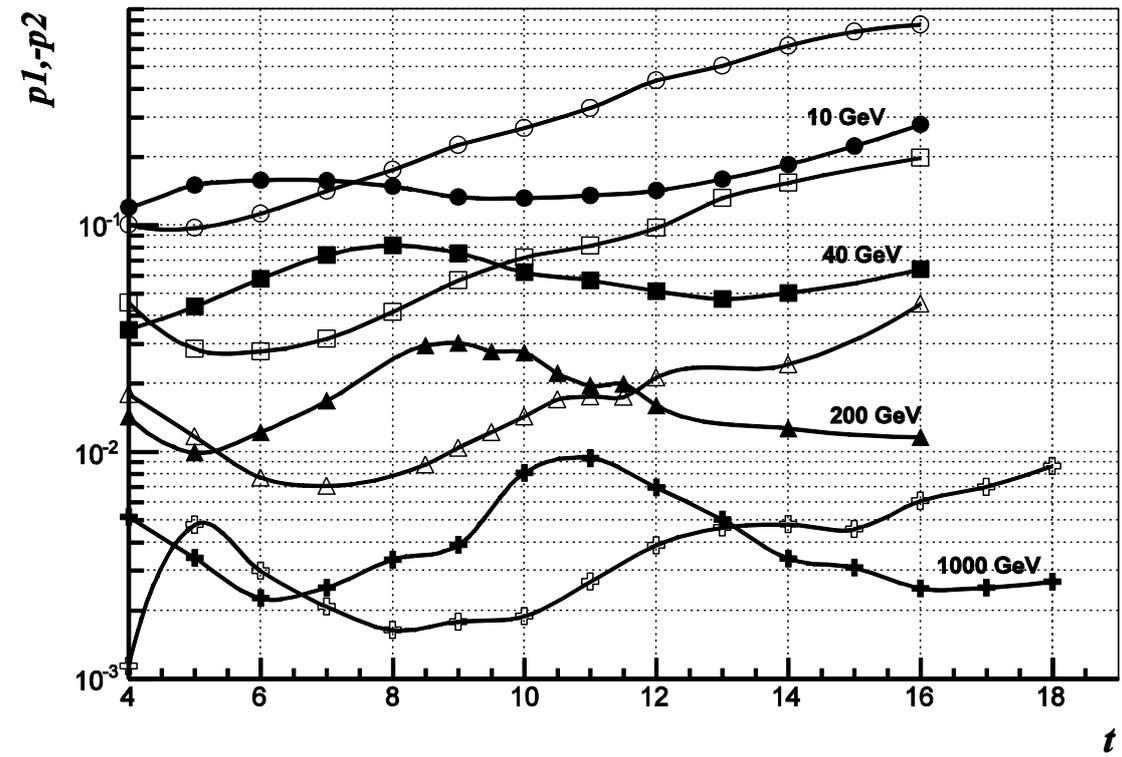

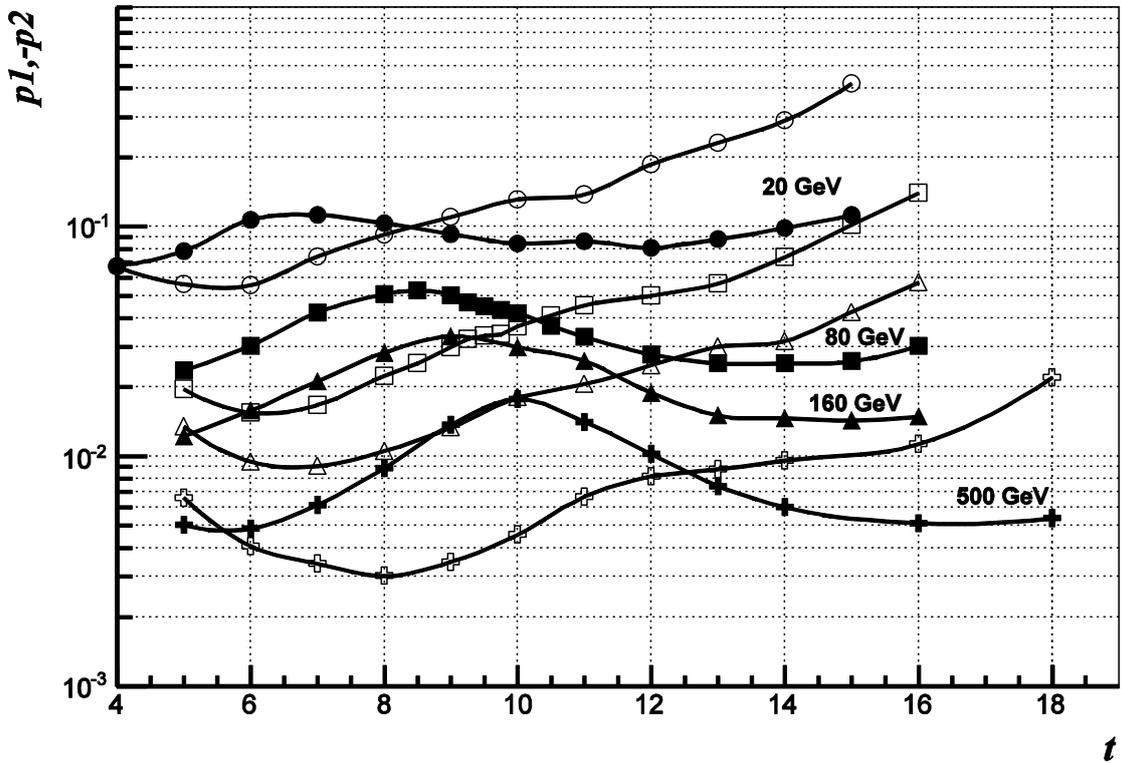

Figure 6. $t$ - dependencies of the $p_1$ (black marks) and $-p_2$ (open marks) parameters for different $E_0$ energies. The smooth curves are drawn using ROOT package.

## 3. RMS/<N> vs $E_0$ and $t$

An important parameter of the particles multiplicity distributions is RMS/<N>. $t$-dependencies of RMS/<N> values for different $E_0$ energies are shown in Fig.7. From Fig.7 it follows that the positions $t_0$ of minimal RMS/<N> values do not coincide with $t_{max}$ where particle flux reaches maximum value, instead close to $t_2$. The $t_0$ and $t_{max}$ energy dependencies are presented in Fig.8. Both follow logarithmic dependence:

$$t_{max} = (3.06 \pm 0.08) + (1.129 \pm 0.016) \ln E_0,$$

$$t_0 = (4.31 \pm 0.06) + (1.177 \pm 0.012) \ln E_0$$

with similar slopes while the $t_0$ constant term is 1.25 r. l. above the constant term for $t_{max}$. The energy dependence of RMS/<N> minimal values is shown in Fig.9. It is fitted by the formula

$$(\text{RMS}/<N>)_{min.} = (0.46 \pm 0.03)/E^{0.49 \pm 0.04} + (0.074 \pm 0.005). \quad (4)$$

The first term in (4) can be interpreted as a stochastic term independent of the place in the converter where the EM shower started to develop. The constant term is mainly due to fluctuations of the shower starting point. Formula (4) is an estimate of the best energy resolution of a simple $e$, $\gamma$ detector, consisting of a high Z converter and a counter of $e^+$, $e^-$ behind it.

## 3. Conclusions

Calculations of the EM showers initiated by 10 to 1000 GeV electrons in lead are performed using GEANT4 to investigate fluctuations of the charged particles fluxes. It is shown that for all studied electron energies and lead depths probability distributions $dP/dN$ of the charged particles multiplicities are well described by the inverse sum of two exponents with three free parameters. At the certain depths $t_1 < t_{max}$ and $t_2 > t_{max}$ $dP/dN$ distributions are symmetric. In the interval from $t_1$ to $t_2$ $dP/dN$ distributions are asymmetric with the tail on the left side of the $dP/dN$ peak and in the region $t > t_2$ the tail appears on the right side of the peak. Such evolution of $dP/dN$ shape can be explained by the late development of some EM showers. Minimal RMS/<N> values of the multiplicity distributions are achieved at the depths close to $t_2$. As follows from formula (4) their $E_0$-dependence obeys a power law. Formula (4) is an estimate of the best energy resolution of a simple $e, \gamma$ detector consisting of a high Z converter and a detector of secondaries $e^+$, $e^-$ behind it.

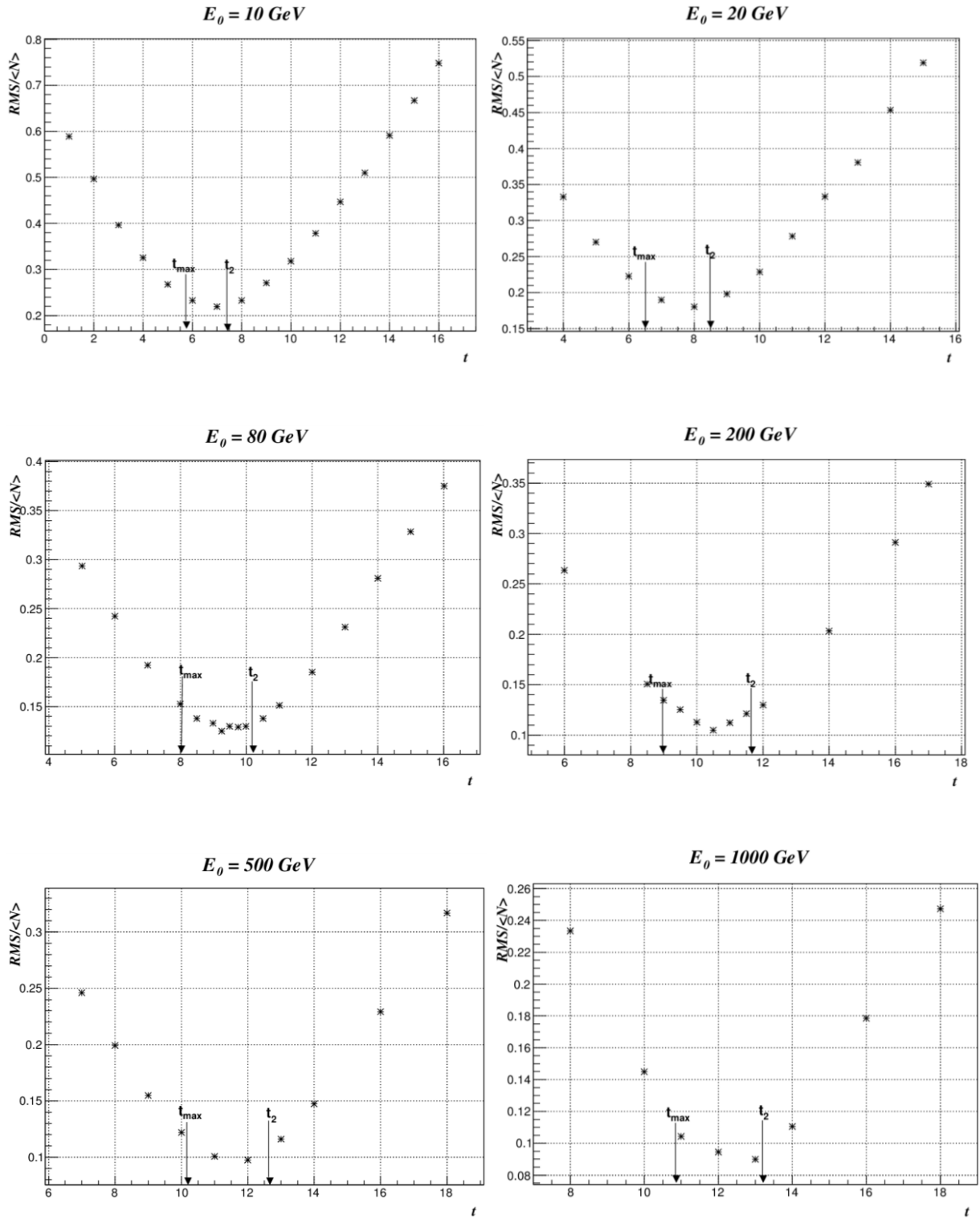

Figure 7. *t*-dependence of RMS/<*N*> for particles multiplicity distributions in the EM showers initiated by 10 to 1000 GeV electrons in lead.

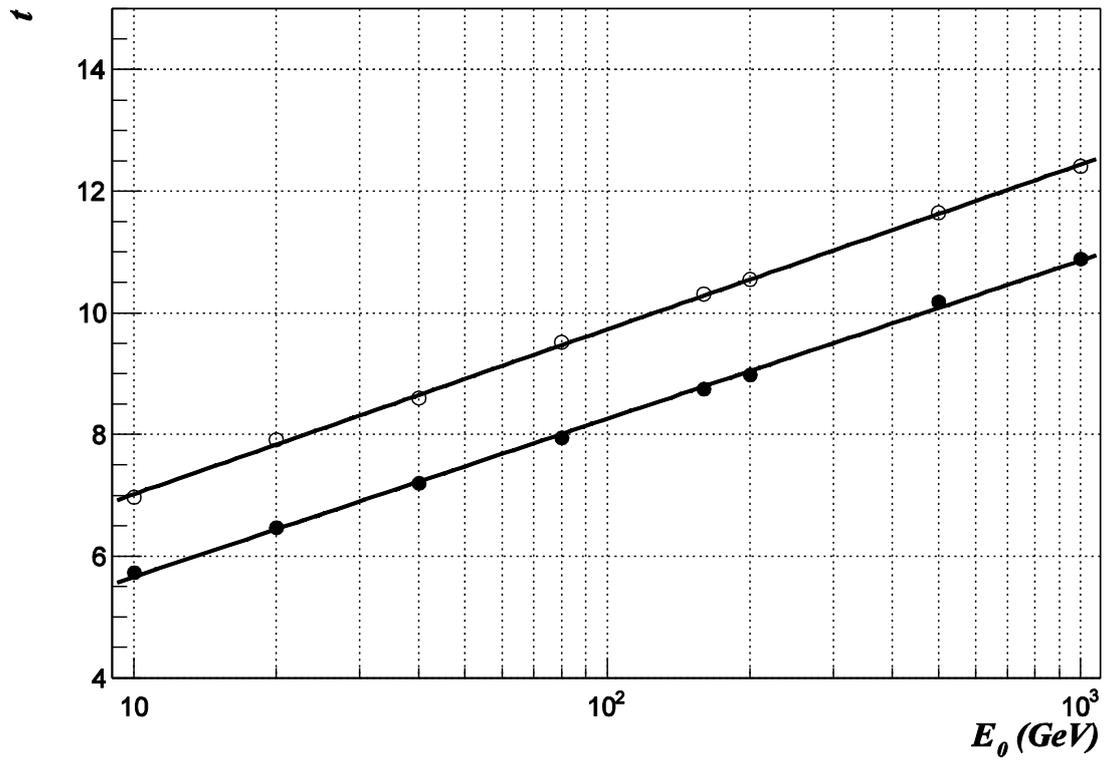

Figure 8. $t_{max}$ (black circles) and $t_0$ (open circles) vs shower energy $E_0$.

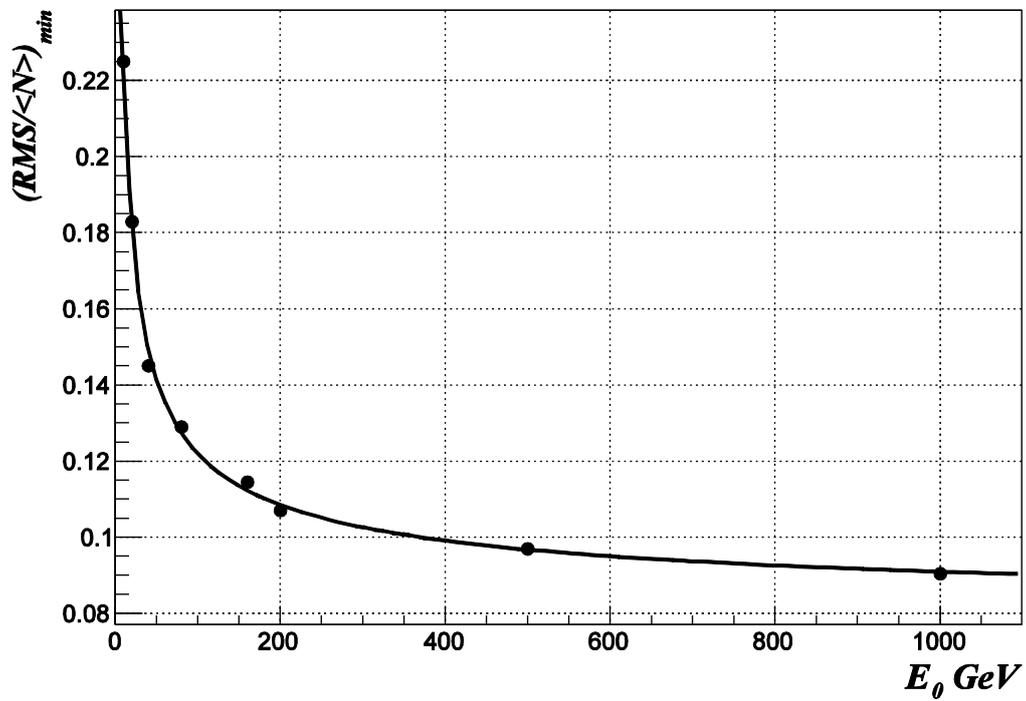

Figure 9. Minimal RMS/$<N>$ values *vs* shower energy $E_0$.

## Acknowledgments

We gratefully acknowledge the help of D.S. Denisov, T.Z. Gurova, and D.A. Stoyanova in preparation of this manuscript. This work was supported in part by the Russian Foundation for Basic Research under grant #17-02-00120.
## References

[1]   A. A. Tyapkin, NIM 85 (1970) 277-278.

[2]   Ts. A. Amatuni, S. P. Denisov, R. N. Krasnokutsky et al., NIM **203** (1982) 179-182.

[3]   Ts. A. Amatuni, Yu. M. Antipov, S. P. Denisov et al., NIM **203** (1982) 183-187.

[4]   G. Apollinari, N. D. Giokaris, K. Goulianos et al., NIM A **324** (1993) 475-481

[5]   Acosta, B. Bylsma, L. S. Durkin et al., NIM A **354** (1995) 296-308.

[6]   S.J. Alvsvaag , O.A. Maeland, A. Klovning et al., NIM A **360** (1995) 219-223.

[7]   K. Byrum , J. Dawson , L. Nodulman et al., NIM A **364** (1995) 144-149.

[8]   S. A. Akimenko, V. I. Belousov, B. V. Chujko et al., NIM A **365** (1995) 92-97.

[9]   J. Grunhaus, S. Kananov, C. Milststene, NIM A **335** (1993) 129-135.

[10]  J. Grunhaus, S. Kananov, C. Milststene, NIM A **354** (1995) 368-375.

[11]  K. Kawagoe, Y. Sugimoto, A. Takeuchi et al., NIM A **487** (2002) 275–290.

[12]  S. Itoh, T. Takeshita, Y. Fujii, F. Kajino et al., NIM A **589** (2008) 370–382.

[13]  A. Ronzhin, S. Los, E. Ramberg et al., NIM A **795** (2015) 288-292.

[14]  S. P. Denisov, V. N. Goryachev, Physics of Atomic Nuclei **81**(2018) No.11, 1-6.

[15]  S. Agostinelli et al, NIM A **506** (2003) 250-303,  http://cern.ch/geant4

[16]  S. P. Denisov, V.N. Goryachev,  Preprint IHEP 15-2018.

[17]  E. Longo, I. Sestili,  NIM **128** (1975) 283.